\documentclass[letterpaper,journal]{IEEEtran}
\usepackage{amsmath,amsfonts}
\usepackage{algorithmic}
\usepackage{algorithm}
\usepackage{array}
\usepackage[caption=false,font=normalsize,labelfont=sf,textfont=sf]{subfig}
\usepackage{textcomp}
\usepackage{stfloats}
\usepackage{url}
\usepackage{verbatim}
\usepackage{graphicx}
\usepackage{hyperref}  
\usepackage{xcolor}
\usepackage{ctable}
\usepackage{array}
\usepackage{caption}
\usepackage{cite}

\begin{document}

\title{Traffic Performance GPT (TP-GPT): Real-Time Data Informed Intelligent ChatBot for Transportation Surveillance and Management}

\author{Bingzhang Wang, Zhiyu (Joey) Cai, Muhammad Monjurul Karim, Chenxi Liu, and Yinhai Wang, {\it Fellow, IEEE}        
\thanks{B. Wang, M.M. Karim, C. Liu, and Y. Wang are with the Department of Civil and Environmental Engineering, University of Washington, Seattle, WA. (Email: bzwang@uw.edu; mmkarim@uw.edu; lcx2017@uw.edu; yinhai@uw.edu)}
\thanks{Z. Cai is with the Department of Civil and Environmental Engineering, University of California, Berkeley, CA. (Email: zhiyu\_cai@berkeley.edu)}
}




\maketitle

\begin{abstract}
The digitization of traffic sensing infrastructure has significantly accumulated an extensive traffic data warehouse, which presents unprecedented challenges for transportation analytics. The complexities associated with querying large-scale multi-table databases require specialized programming expertise and labor-intensive development. Additionally, traditional analysis methods have focused mainly on numerical data, often neglecting the semantic aspects that could enhance interpretability and understanding. Furthermore, real-time traffic data access is typically limited due to privacy concerns. To bridge this gap, the integration of Large Language Models (LLMs) into the domain of traffic management presents a transformative approach to addressing the complexities and challenges inherent in modern transportation systems. This paper proposes an intelligent online chatbot, TP-GPT, for efficient customized transportation surveillance and management empowered by a large real-time traffic database. The innovative framework leverages contextual and generative intelligence of language models to generate accurate SQL queries and natural language interpretations by employing transportation-specialized prompts, Chain-of-Thought prompting, few-shot learning, multi-agent collaboration strategy, and chat memory. Experimental study demonstrates that our approach outperforms state-of-the-art baselines such as GPT-4 and PaLM 2 on a challenging traffic-analysis benchmark TransQuery. TP-GPT would aid researchers and practitioners in real-time transportation surveillance and management in a privacy-preserving, equitable, and customizable manner.
\end{abstract}

\begin{IEEEkeywords}
Transportation Analytics, Pre-Trained Large Language Models, SQL Database.
\end{IEEEkeywords}

\section{Introduction}

\IEEEPARstart{W}{ith} the rapid digitization of sensing infrastructure, an immense volume of traffic data is being collected at an ever-increasing rate. This presents both opportunities and challenges. On one hand, these unprecedented data resources hold the promise of propelling advancements in traffic analysis, making it more accurate and reliable; On the other hand, the unpredictable accumulation of data poses significant challenges for the development of sophisticated traffic analysis techniques. Firstly, real-time traffic sensing data is typically stored in large-scale, multi-table databases. These databases are incredibly large and have complex relationships between different data points. Querying these database can be labor-intensive and time-consuming, leading to significant latency. To properly manipulate databases even requires not only a deep preliminary understanding of context but also specialized expertise in database programming. Secondly, traditional methods of traffic analysis have primarily focused on the numerical aspects of the data, using statistical methods and machine learning techniques. This overlooks the semantic and interpretability attributes of the data. As a result, the exploration of traffic datasets has been limited to numerical imputations and prescriptive visualization, neglecting their inherent natural-language significance. Thirdly, access to real-time traffic data is typically restricted to authorized entities such as government agencies and academic institutions, due to privacy concerns, making it inaccessible for direct data handling by the general public. There is a pressing need for an intermediate framework that processes and interprets data for practitioners in a privacy-preserving manner, which has the potential to not only facilitate trip planning and policy-making but also improve traffic analysis by making it more accessible, efficient, and equitable.

In recent years Large Language Models (LLMs) have emerged as a groundbreaking development in the field of artificial intelligence (AI), demonstrating unparalleled capabilities in understanding, and interpreting real-world scenarios in human-like ways\cite{chang_survey_2023}. These models, powered by advanced neural network architectures, have found applications across a wide range of domains, from natural language processing and machine translation to content generation and beyond. They offer unprecedented capabilities in data interpretation and decision-making. 

%

{LLMs have proven their versatility in different fields, including education, healthcare, software engineering\cite{fraiwan_review_2023}, among many others. While still a new area of exploration, LLMs show promise for analyzing traffic data. Their general ability to adapt to different tasks makes them attractive for various intelligent transportation applications, including traffic signal control \cite{lai_large_2023}, and accident risk assessment \cite{wang_accidentgpt_2023}. However, effectively using LLMs in these specialized areas requires them to have a deep understanding of the specific domain \cite{ge_openagi_2024}. This necessitates the integration of domain-specific databases with LLMs to foster data-driven reasoning capabilities. Recent research has explored leveraging LLMs for database operations. Frameworks like LangChain facilitate efficient interaction with LLMs \cite{topsakal_creating_2023}, while models like DB-GPT are fine-tuned with domain-specific database knowledge\cite{zhou2024db}, enabling the translation of textual semantics into database queries (text-to-SQL). These advancements offer exciting possibilities for advanced analysis of large-scale traffic data. However, existing models primarily target general-purpose database tasks, neglecting the unique knowledge domain of traffic data. This omission of traffic-specific background knowledge during training may lead to diminished query result accuracy and limit their applicability in traffic research.




Despite the proliferation of LLM applications across various domains, their potential in traffic data analysis, especially within databases, remains an open research topic. Furthermore, there is a pressing need for a user-friendly intelligent platform that can effectively communicate and analyze real-time data. The paper aims to delve into this gap by proposing an intelligent transportation analytics system that applies pre-trained large language models to complex traffic data analysis. It leverages their capability to generate accurate SQL queries and natural language interpretations based on contextual awareness, demonstrating the extensive pre-trained knowledge of LLMs and their proficiency in adapting to the transportation domain.


The contributions of this paper are listed below:
\begin{enumerate}
    \item An intelligent online chatbot, Traffic Performance GPT (TP-GPT), is proposed for efficient personalized transportation analysis and management leveraging the support of big real-time traffic data. To the best of our knowledge, TP-GPT is the first real-time traffic analysis chatbot empowered by LLMs to be proposed.
    \item Leveraging contextual and generative intelligence of the Generative Pre-trained Transformer (GPT), an innovative framework is constructed to serve as a connection between public users and authorized data resources in a privacy-preserving, equitable, customizable way.
    \item The developed chatbot is able to generate reliable, responsive and accurate traffic analysis and management responses to input questions, by integrating designed prompts, few-shot learning module, multi-agent collaboration strategy and conversation memory module. The proposed method outperforms existing general-purpose LLMs regarding traffic-domain analysis performance.
\end{enumerate}{}{}

\section{Related Work}




\subsection{Traffic Management with LLMs}

Traffic management for road networks especially in urban area encounters complex background knowledge, and real-time demands. Previous research usually relied on traditional hardware infrastructure, such as surveillance cameras and loop sensors to observe traffic flow. Researchers have employed both micro and macro methods \cite{wang_advanced_2023} based on theories and data from traditional methods to simulate traffic flow, promoting the advancement of theoretical research as well as management strategies. 

Recent research addresses challenges of traffic more from data-driven perspectives with the emerging techniques of mobility and traffic data collection and analysis. There has been a shift towards leveraging deep learning methods and LLMs to provide new solutions to traffic control and management. Traffic control at intersections\cite{lai_large_2023} emphasizes traffic efficiency and collision prevention. Other scenarios such as ring and bottleneck scenarios have also been studied \cite{villarreal_can_2023} combining with robotic vehicles and human drivers.

Other research expanded the scope of specific scenarios, such as \cite{zhang_trafficgpt_2024} combines LLMs to assist human users in traffic network analysis and further decision-making. There are more research explored into forecasting tasks. A traffic prediction model proposed by researchers in \cite{ren_tpllm_2024} shows capacity to tackle flow prediction tasks under full-sample and few-shot historical data scenarios. By encoding the data in a specific way\cite{sun_test_2023}, LLMs can be widely used to conduct research on time series and spatio-temporal data\cite{liu_spatial-temporal_2024}, and their potential to analyze data of such modalities is being further explored, including univariate time series forecasting\cite{rasul_lag-llama_2023}, multivariate time-series data\cite{jin_time-llm_2023}.

One other significant challenge is model's capability to understand contexts in transportation systems. Authors in \cite{wang_transgpt_2024} manually collected multi-modal traffic data sets,such as text and traffic signs alignment for model training, while authors in \cite{tang_synergizing_2024} focused on users' itinerary planning demands. Among a variety of integration of Large Language Models into traffic management, this paper showcases a promising avenue for enhancing real-time traffic analysis and management in order to benefit practitioner, traveler and researcher.

\subsection{LLMs in Database SQL Query} 
LLMs exhibit remarkable capabilities in generating reliable and accurate SQL queries, advancing database interaction efficiency. A recent study introduces BIRD\cite{li_can_2023}, a benchmark for large-scale database text-to-SQL tasks highlighting challenges such as content quality, external knowledge integration, and SQL efficiency. Their inherent contextual intelligence enables them to understand data origins and relationships, thus enhancing the performance of data queries and analyses.  Many scholars use SQL agent to  enhance database-related tasks, such as \cite{hong2024knowledge} employs a Data Expert LLM (DELLM) to provide knowledge for text-to-SQL models to generate accurate queries. Authors in \cite{cai_star_2022} enhance text-to-SQL parsing in multi-turn conversations by leveraging contextual information from the dialogue history. Research \cite{li2024using} utilizes LLMs to automatically generate test cases for selecting the most accurate SQL query from a set of candidates. 

Databases can be queried more effectively by leveraging LLMs, ensuring that target data is extracted with a high degree of precision. This prowess underscores the potential of LLMs in database management and analysis. Current research mostly uses agents for task processing from the perspective of context. The disadvantage of this is that LLM's ability to perform SQL queries is relatively static and requires a large number of iterations to correct the model.





\section{Methodology}

\subsection{Problem Statement} \label{problem statement}

A substantial magnitude of traffic intensity is being observed in modern mobility. As reported by the Washington State Department of Transportation, the annual Vehicle Miles of Travel (VMT) on the state highway of King County reached a total of 8,534 million in the year 2022. Meanwhile, the wide-range deployment of traffic sensing systems has employed tons of real-time and historical data. The unpredictable, fast-changing traffic patterns underlying the enormous numerical data pose an ever-challenging task for efficient and effective traffic analysis and management. To tackle this problem, we incorporate a large-scale network-wide mobility database hosted by STAR Lab\cite{tpswebsite}, which integrates the real-time data resources of traffic counts (i.e., speed, volume, occupancy) collected from more than 8,000 inductive loop detectors, as well as the route segment-wide Traffic Performance Score (TPS) calculated using Equation \ref{eq:def_of_tps} based on loop data\cite{cui2020traffic}. 

\begin{equation}
TPS_t = \frac{\sum_{i=1}^n V_t^i \cdot Q_t^i \cdot L^i}{\sum_{i=1}^n V_f \cdot Q_t^i \cdot L^i} \times 100\%
\label{eq:def_of_tps}
\end{equation}
where $V_t^i$ and $Q_t^i$ represent traffic speed and volume of road segment $i$ at time $t$. $L^i$ is road segment length covered by the $i$-th detector. $V_f$ is free flow speed. Therefore, TPS is a value ranging from 0\% to 100\% where 0\% is the worst traffic condition and 100\% is the best.

These loop detectors are deployed on freeways, including I-5, I-90, I-99, I-167, I-405, and SR-520, in the Greater Seattle Area, WA, as shown in Figure \ref{fig:loop}. Data have been collected online in one-minute intervals from 2020 to the present, and it has accumulated around 1.89 Terabytes (TB) in the data warehouse. There are a total of 6 tables in the database, with the details shown in Table \ref{tab:database}.

\begin{figure}[h]
    \centering
    \includegraphics[width=0.8\columnwidth]{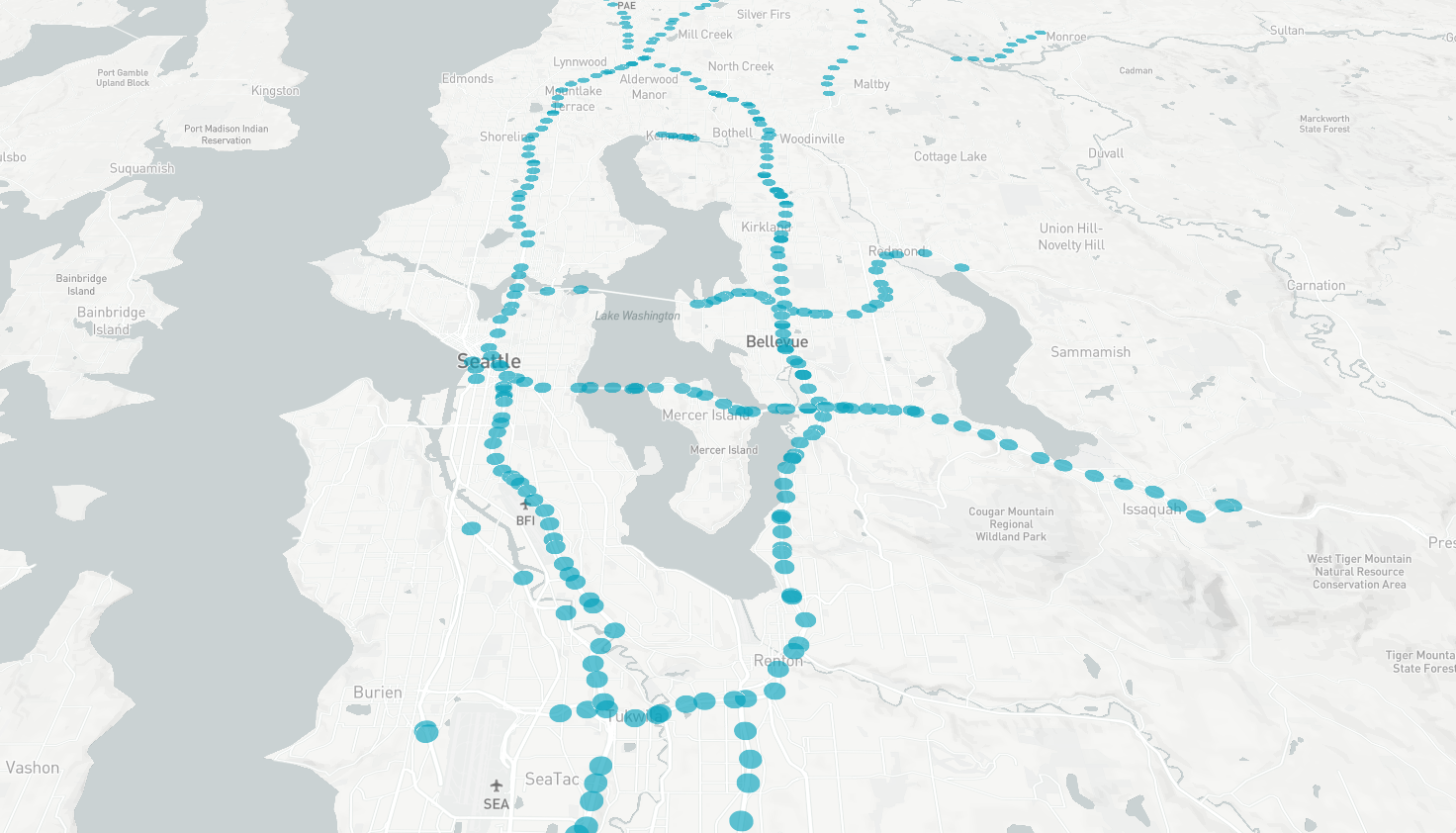}
    \caption{Inductive loop Detector locations shown as blue points}
    \label{fig:loop}
\end{figure}

This level of database has great potential for insightful network-wide analysis of real-time traffic performance. However, vast volume of data poses great challenges for practitioners' and researchers’ inspection and management of data. Firstly, exploring the data warehouse necessitates participants' deep understanding of database setup details (i.e., tables, columns, data types, relations, and so on), as well as expertise in database programming especially processing large-scale databases. Even though, manually programming various queries and then executing commands for real-time information in such a huge database is particularly time-consuming and labor-intensive, making it almost a mission impossible. Second, interpreting from numeric data results to human-language traffic advisory or impactful analysis results needs in-depth professional transportation-domain knowledge and a wide range of historical knowledge base support such as urban planning, and social events. Third, the lack of direct access to the database for security and privacy concerns limits the potential participants’ exploration, hindering flexible investigation. 

\newcolumntype{P}[1]{>{\centering\arraybackslash}p{#1}}

\begin{table}[h]
    \caption{Network-wide traffic database introduction}
    \label{tab:database}
    \centering
    \resizebox{\columnwidth}{!}{
    \begin{tabular}{c c P{0.5\columnwidth}}
        \hline
        \specialrule{.2em}{.1em}{.1em}
        \textbf{Table Name} & \textbf{Columns} & \textbf{Description}\\
        \hline
        \specialrule{.2em}{.1em}{.1em}
        \textit{dbo.cabinets} & 17 & Loop detector details of unit name, coordinate, route, milepost, and direction.\\
        \hline
        \textit{dbo.cabinfo} & 6 & District location of loop detectors sorted by cabinet station ID. \\
        \hline
        \textit{dbo.MinuteDataNW} & 6 & One-minute traffic speed, volume, occupancy data in Washington Northwest sorted by loop detector ID and timestamp.\\
        \hline
        \textit{dbo.Segments} & 6 & Road segment definition with corresponding location information.\\
        \hline
        \textit{dbo.SegmentTrafficIndex} & 8 & Segment-based traffic performance data on general-purpose lanes and carpool lanes, including speed, volume, and TPS.\\
        \hline
        \textit{dbo.TrafficIndex} & 9 & Statistical traffic performance data for each defined road segment. \\
        \hline
        \specialrule{.2em}{.1em}{.1em}
    \end{tabular}
    }
\end{table}

In this context, we propose an intelligent traffic performance chatbot TP-GPT for real-time transportation surveillance and management in a privacy-preserving way. While, technical problems need to be tackled for comprehensive development: 
1) ChatGPT does not have preliminary contextual knowledge of the database setup, structures and content. Even though ChatGPT has strong abilities in programming and inference, direct deployment of the language model to database analysis without prerequisites may result in hallucination that blind actions irrelevant to the raised question would possibly be taken. Thus, the compilation of appropriate input prompts is essential for employing ChatGPT. 2) ChatGPT has a reliable performance of executing sequential commands activated by human's input. However, the lack of monitoring the response relevance and query correctness regarding to the user's question may generate unexpected, unreliable answers. It is necessary to incorporate an intelligent autonomous streamline with self-management capability for quality control of answer generation. Specifically, iterative communications between ChatGPT and the database in the revision circulation is needed to keep modifying responses until certain standard is reached. This process necessitates the participation of multiple intelligent virtual agents of different roles, such as database query engineer, quality manager, consultant and more. 3) The demonstration of a set of typically asked questions with their corresponding answers are critical for ChatGPT to specify the work scope, furthermore, to improve the effectiveness and efficiency of target response. 4) In order for users to be able to consecutively interact with the chatbot, a chatting memory function is needed so that the chatbot has the historical memory to support following generative process.

\subsection{System Design}

\begin{figure*}[!t]
\centering
\includegraphics[width=\textwidth]{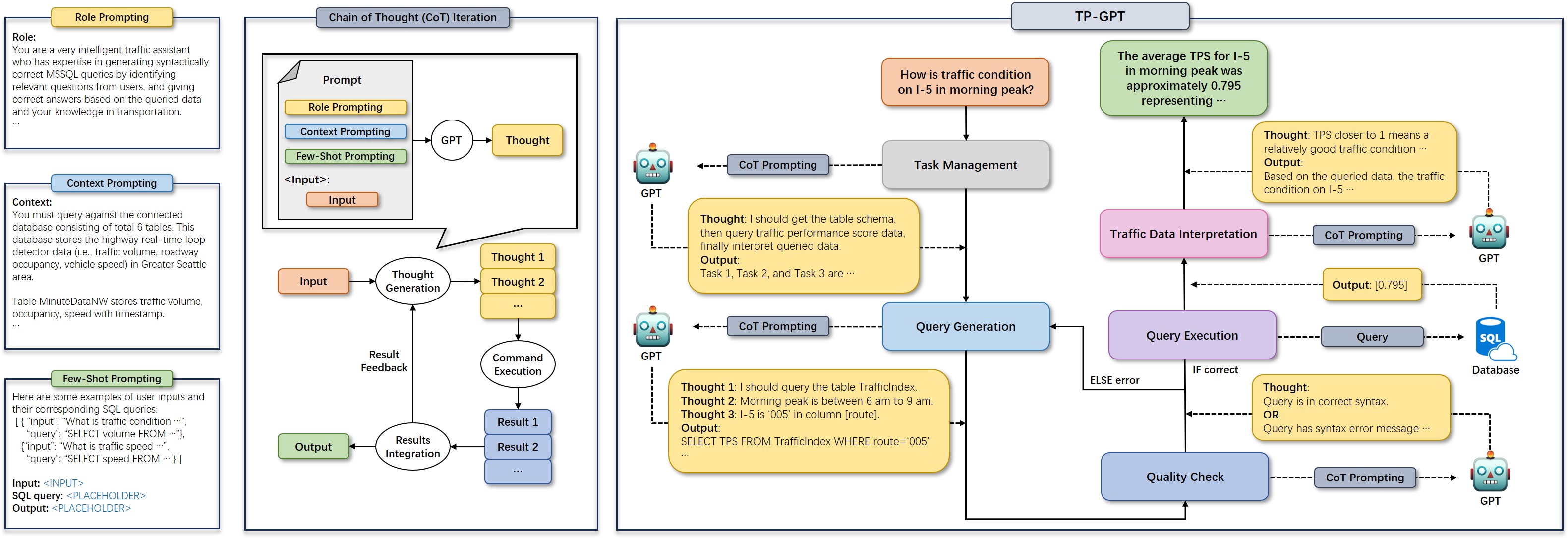}
\caption{Overall system architecture design demonstrating traffic analysis pipeline (right), example input prompts (left), and incorporated Chain of Thought iterative prompting (center)}
\label{fig:system}
\end{figure*}

TP-GPT aims to answer users' questions by integrating GPT with real-time traffic SQL database. Directly applying generative model to the database has potential to cause several concerns such as reasoning hallucination, answer accuracy, and performance reliability. To overcome these obstacles, the proposed framework incorporates multiple modules as a pipeline to process and interpret data, as shown in Figure \ref{fig:system}. Specifically, once users' questions are input into the ChatBot, five steps are executed, each of which is empowered by GPT's contextual and generative intelligence, to generate the final answer: 1) Task management: Since the fixed sequence of implementation steps is not adaptable to resolving different questions, a customized execution plan is produced based on the input. Meanwhile, each execution step is determined and managed flexibly by GPT's decision-making capability. For example, if the question is relevant to searching the traffic data, database query generation and execution will be included in the plan; Otherwise, the chat function will be activated to interact with users based on the general knowledge base. 2) Query generation: A query to extract relevant data in the database is generated by understanding the input question. Constraints on query statement specifics such as programming syntax and quantity of queried data are applied to the process. 3) Quality check: An initial syntax check is performed on the generated query to inspect if it is beyond the designated constraints like too many data instances are queried. 4) Query execution: The checked query is executed in the SQL database. While, an inspection of the execution log is conducted to debug any existing errors. 5) Data interpretation: The queried data combined with the user's question is analyzed and interpreted into a traffic advisory report incorporating GPT's wide range of transportation knowledge base.

\subsection{Input Prompt Generation}
Prompt engineering is essential for LLMs to master preliminary contextual information before a practical question is fed. Such integration can not only enable language models to avoid the unnecessary repeating process of reasoning users' desired responses in each run when the contexts (i.e., database setup, question background) are constant, but also boost the ChatBot performance in the transportation domain. In the proposed framework, the iterative interface with GPT is achieved through the Chain of Thought (CoT) prompting \cite{NEURIPS2022_CoT}, which iteratively reasons a series of intermediate steps before the final answer is generated. The conceptual workflow of CoT iteration is shown in Figure \ref{fig:system}. The designed prompts cover comprehensive descriptions in these aspects: 1) Instructions on what role the model is expected to play in performing the tasks and how to generate the object responses; 2) In-depth descriptions of database setup with an emphasis on table relations, column explanation, and data significance. 3) Domain knowledge in transportation scenarios, especially those missing from GPT's commonsense, such as the definition of Traffic Performance Score and its measurement scale. 4) Designated output format that is desired from GPT at the current step. An example prompt input is shown in Figure \ref{fig:system}.

\subsection{Multi-Agent Strategy}
GPT achieves superior performance in solving individual tasks following instructional logic in a dialogue. However, highly sophisticated problems in real-world traffic analysis are challenging to tackle by directly adopting GPT's general intelligence. Here we introduce the multi-agent strategy in transportation scenarios, an innovative collaborative workflow incorporating GPT as different agents to work as a team, which simulates the human's real teamwork logistics in a research lab or industrial company. The overall objective is to decompose the original complex tasks into sub-tasks of different scopes and then assign each to a GPT agent expertized in the designated domain to solve by a streamlined, collaborative workflow. The interaction between agents is enabled and achieved by incorporating a public scratchpad for the agent team to track the updated progress. Agents iteratively communicate with the GPT based on the CoT prompting dialogue in JSON format, illustrated in the previous section, to seek advisory and thoughts on task solutions and then formulate the execution plan.

\begin{figure}[h]
\centering
\includegraphics[width=\columnwidth]{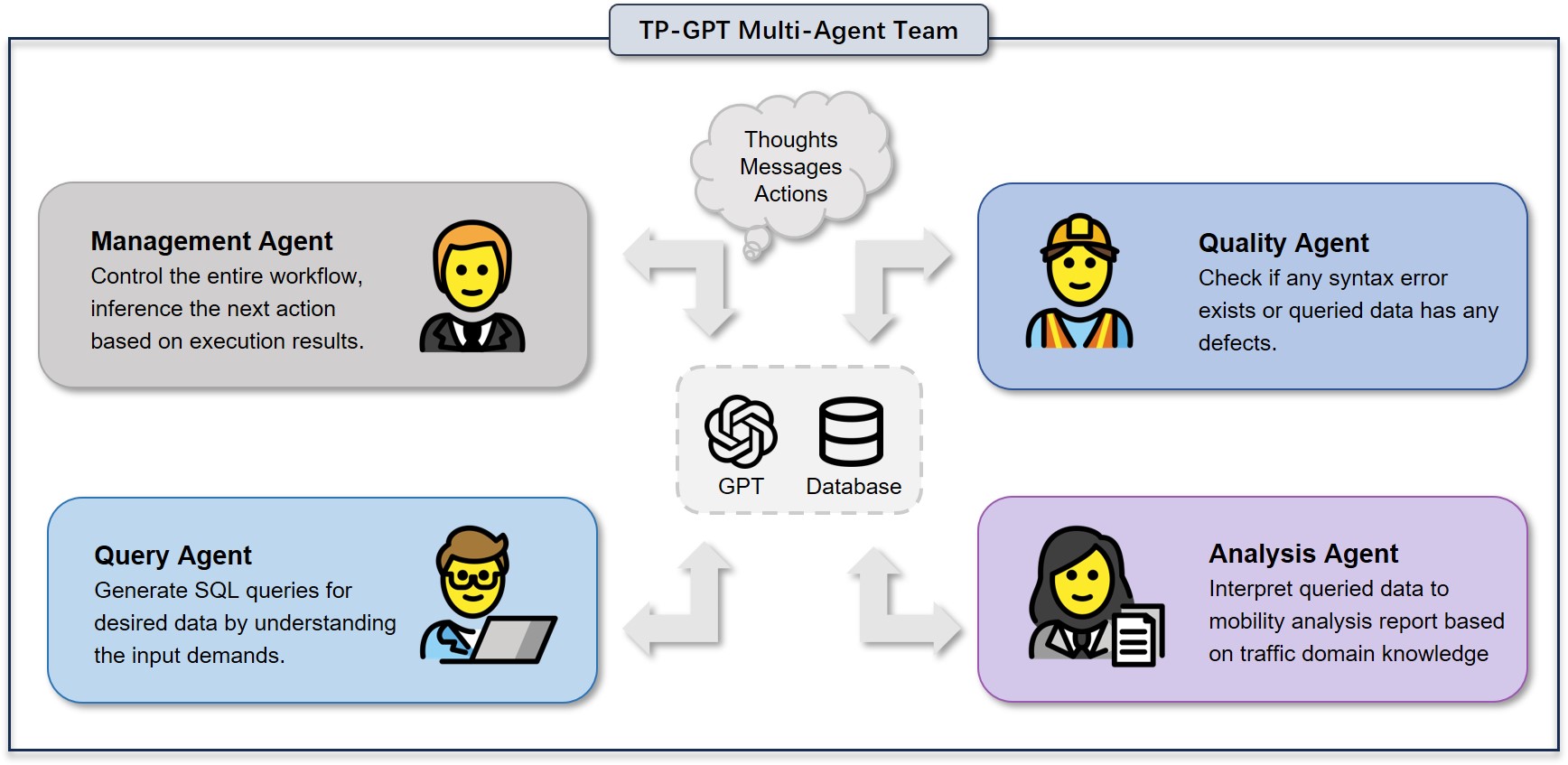}
\caption{TP-GPT multi-agent collaboration framework for traffic data analysis}
\label{fig:multiagent}
\end{figure}

The joint collaborative framework includes four virtual agents, shown in Figure \ref{fig:multiagent}, each of which conducts the designated scope of tasks corresponding to a teamwork role. In detail, the project manager controls the entire team's workflow by reasoning the next-step action based on the execution results from the last step; SQL engineer generates synthetically correct queries for the desired data through contextual understanding of the users' input demand; Quality analyst checks if a syntax error exists in the generated query or the obtained data is not reasonably acceptable, then necessary inspections into backend database execution log are implemented to figure out the potential cause to the error in order to correct the query; Finally, data analyst interprets the queried data to mobility analysis report based on pre-trained transportation domain knowledge, to answer the user's demanded information in a natural-language illustrative method with real-time data and detailed explanation involved. The multi-agent collaborative strategy significantly improves the overall performance of the ChatBot in terms of accuracy, reliability, interpretability, and flexibility.  

\subsection{Few-shot Learning Prompting}
Although LLMs exhibit impressive zero-shot capabilities for general problems, they struggle to address complex problems in the transportation domain using only a zero-shot approach. Few-shot prompting can be used as a technique to facilitate in-context learning by providing examples that guide the model toward better performance. The demonstrations in the prompt contain exemplar user questions with their ideal queries crafted by developers that the model is expected to follow to generate subsequent responses. In each attempt to resolve the input question, the model can refer to several of the most relevant examples, which not only enables the model to master the contextual knowledge required to answer similar questions accurately but also improves response efficiency by significantly decreasing reasoning time and the risk of causing errors. To further enhance response reliability leveraging few-shot learning, an example repository covering multiple transportation scenarios is manually crafted, including real-time traffic advisory, historical data statistical analysis, travel emission inquiry, lane-based traffic performance inspection, and so on, for providing comprehensive transportation evaluations. In implementations, the input question is converted to text embedding to search the designated number of most similar questions in the example repository. Then, these question-query example pairs are used to populate the generated prompt for model inference. A typical few-shot prompt is displayed in Figure \ref{fig:system}. This approach avoids offering irrelevant examples to target work scope, thereby increasing the few-shot efficiency and accuracy, as well as reducing the possibility of non-objective examples misleading the response. The integration of few-shot learning remarkably enhances ChatBot performance, especially when questions that are typically asked fall into the pre-defined scope. This also provides the potential for further extension of objective response domain by simply adding more exemplar use cases.  

\subsection{Chat Memory}
To enhance interactivity, the ChatBot should retain a memory of past conversations, allowing it to infer and tailor future responses based on previous interactions. In most use cases, if the initial generated response does not fulfill users' expectations, normally subsequent questions will follow, assuming chat history is held. Thus, the ChatBot is integrated with the chat memory module by saving dialogue history in each conversation session. Specifically, only the part of chat records relevant to the current input question is retrieved from memory for the model's reference in response, which prevents from reading the whole lengthy dialogue list. Afterward, the generated answer, along with the question, is written into memory. This simple yet effective implementation advances the ChatBot towards more efficient and intelligent interaction experience. 

\section{Experiments}
An online ChatBot with interactive web interface has been developed using Streamlit framework in Python. An example conversation between user and TP-GPT chatbot is visualized in Figure \ref{fig:interface}. Furthermore, experimental study has been conducted leveraging the ChatBot interface to evaluate and compare performance of TP-GPT with baseline models. 

\begin{figure}[h]
    \centering
    \includegraphics[width=\columnwidth]{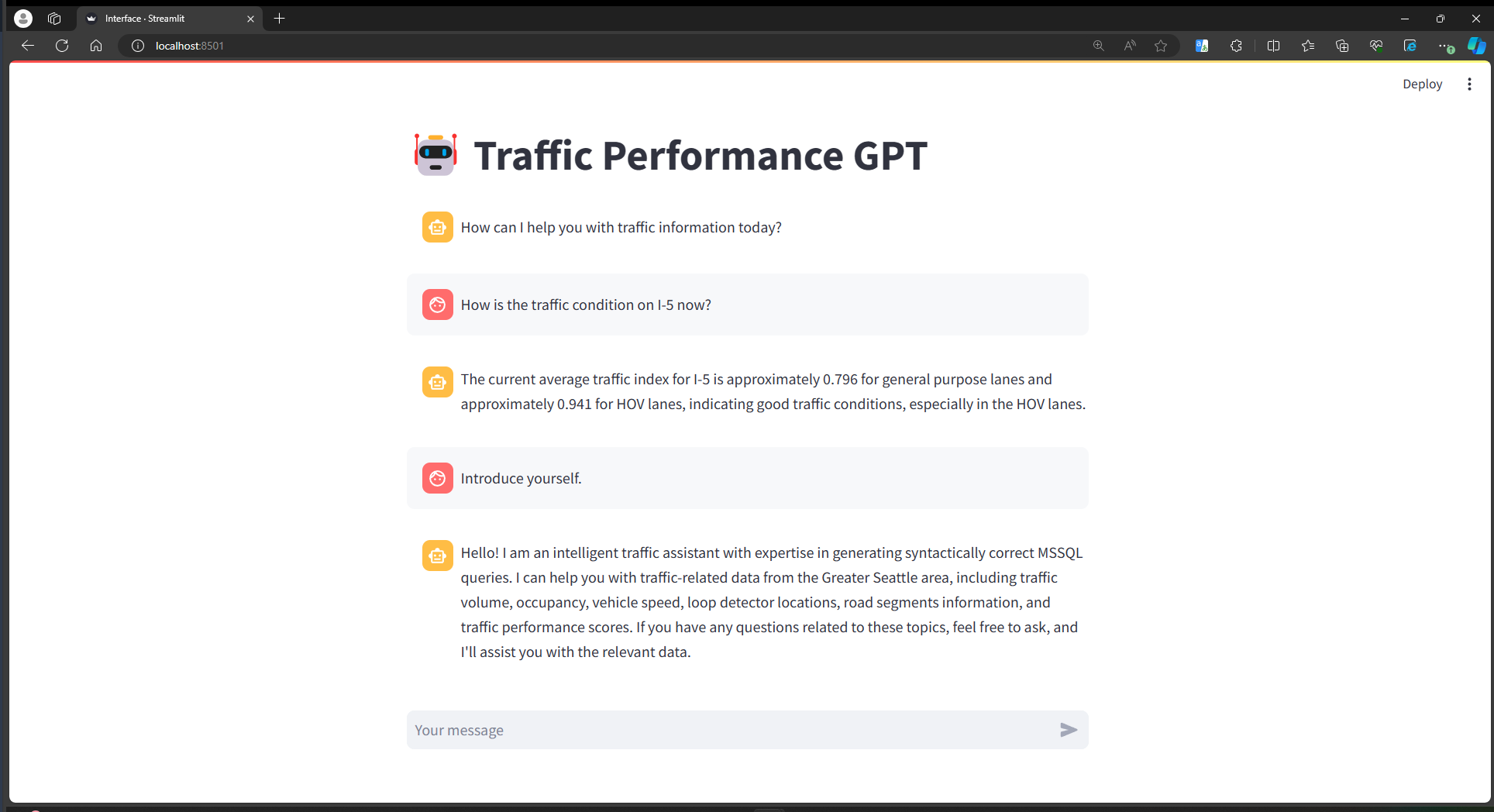}
    \caption{Online web application interface of TP-GPT ChatBot}
    \label{fig:interface}
\end{figure}

\subsection{Experimental Setup}\label{Experimental Setup}
To demonstrate the model's performance in solving transportation-specific domain problems, we introduce our self-generated challenging traffic analysis benchmark named TransQuery based on the mobility database as mentioned in Section \ref{problem statement}. The benchmark includes 50 manually designed complicated transportation surveillance and management tasks in real-world scenarios. Each instance is an input-output pair consisting of a question and its ground-truth response (i.e., SQL query, target data, and natural language answer). These tasks cover a wide scope of traffic analysis challenges including but not limited to: spatial-temporal traffic condition inquiry, traffic pattern recognition, peak traffic behavior analysis, real-time lane-based trip advisory, traffic counting statistics, and vehicle emission evaluation. Some of the example problems are listed in the Table \ref{tab:question}. To demonstrate the validity of employing TP-GPT framework, we use TransQuery benchmark to compare our method with GPT-4 Turbo\cite{gpt4website}, PaLM 2\cite{anil2023palm}, and SQLCoder\cite{sqlcoderwebsite} which are state-of-the-art baseline LLMs having expertise in the database query domain. Standard prompting containing fundamental database information and contextual knowledge is integrated into baseline models.

\begin{table}[h]
\caption{Sample questions in TransQuery\label{tab:question}}
\centering
\renewcommand{\arraystretch}{1.5}
\resizebox{\columnwidth}{!}{
\begin{tabular}{c}
\hline
\specialrule{.2em}{.1em}{.1em}
\textbf{Input questions}\\
\hline
\specialrule{.2em}{.1em}{.1em}
Show me the most recent loop data on SR 99 Northbound sorted by locations.\\
\hline
Compare traffic performance of I-5 between Monday and Sunday in last month.\\
\hline
What was the traffic condition of SR-520 during today's evening peak?\\
\hline
Should I use HOV or general purpose lane on I-5 now? \\
\hline
How many cars are on each segment of I-405 on average now?\\
\hline
 What is the difference in greenhouse gas emissions between weekdays and weekends in last month?\\
\hline
\specialrule{.2em}{.1em}{.1em}
\end{tabular}
}
\end{table}

To statistically compare the performance of each model on TransQuery, the execution result of each instance of the benchmark is evaluated by the metrics with corresponding rate scores: \textbf{Non-functional}: SQL query is non-executable due to existing errors, with rate score of 0; \textbf{Runnable but imperfect}: SQL query can be executed successfully but not perfectly answers the input question, with rate score of 1; \textbf{Flawless}: correct data can be queried, and interpretation of data can properly answer the input question, with rate score of 2. In this way, an average performance score $S$ is calculated for each model using the Equation \ref{eq:def_of_metrics}:

\begin{equation}
S = \frac{\sum_i n_i s_i}{s_{max} \sum_i n_i}
\label{eq:def_of_metrics}
\end{equation}
where $i$ represents evaluation categories (e.g., non-functional), $n$ represents the number of experimental instances classified to each category, and $s$ represents their corresponding rate scores.

\subsection{Performance Comparison}
In experiments, three baseline models are compared to TP-GPT using TransQuery benchmark that has 50 complex transportation problems. As shown in Table \ref{tab:summary}, TP-GPT significantly outperforms other baseline models on TransQuery with an average performance score of 0.87,  surpassing GPT-4 Turbo of 0.57, PaLM 2 of 0.34, and SQLCoder of 0.16. Specifically, out of total 50 questions, TP-GPT generates 40 flawless responses accounting for 80\% of the whole dataset, 7 runnable queries but imperfect responses, and only 3 non-functional queries. Whereas, the state-of-the-art LLM GPT-4 Turbo only generates 22 flawless responses accounting for 44\% of the whole dataset, even with superior pre-trained reasoning and contextual capability. PaLM 2 and SQLCoder barely generate 15 and 7 flawless responses accounting for the percentage of 30\% and 14\% respectively. 

\begin{table}[h]
\caption{Performance comparison of TP-GPT with baseline models on TransQuery traffic-analysis benchmark}
\label{tab:summary}
\centering
\renewcommand{\arraystretch}{1.5}
\resizebox{\columnwidth}{!}{
\begin{tabular}{c c c c c}
\hline
\specialrule{.2em}{.1em}{.1em}
Model & \multicolumn{3}{c}{Response Metrics} & Avg. Score\\
& Non-functional & Runnable but imperfect & Flawless \\
\hline
\specialrule{.2em}{.1em}{.1em}
PaLM 2 & 31 (62\%) & 4 (8\%) & 15 (30\%) & 0.34\\
\hline
SQLCoder & 41 (82\%) & 2 (4\%) & 7 (14\%) & 0.16\\
\hline
GPT-4 Turbo & 15 (30\%)& 13 (26\%)& 22 (44\%) & 0.57\\
\hline
\textbf{TP-GPT} & 3 (6\%) & 7 (14\%) & 40 (80\%) & 0.87\\
\hline
\specialrule{.2em}{.1em}{.1em}
\end{tabular}
}
\end{table}

This result not only demonstrates the complicacy and challenge of TransQuery benchmark, but also indicates TP-GPT's superior performance in resolving transportation scenario problems and its effectiveness when GPT is integrated with our developed ChatBot system. A detailed comparison between the models is narrated in the following. 

\begin{itemize}
  \item \textbf{PaLM 2}: The most capable PaLM 2 model in code generation named Bison, is tested as one of the baselines due to its compatibility with SQL query generation. This model has the fastest response time for its relatively small model size. However, the performance struggles in dealing with complex tasks requiring overloaded tokens that frequently exceed the limit. For example, the question "\textit{Compare traffic performance of I-5 between Monday and Sunday in last month.}" requires querying a vast amount of minute-by-minute historical data in one month to summarize the traffic performance. This data volume cannot be processed by Bison causing the token limit error.
  \item \textbf{SQLCoder}: Defog's SQLCoder-34B is implemented for text-to-SQL performance comparison as one of baselines. The model is constrained by its model size achieving the lowest solve rate. It could have been trained using the dataset that consists of MySQL-syntax samples other than MSSQL dialect, which is potentially the major factor of disadvantageous performance. However, SQLCoder provides great possibilities of future fine tuning to adapt to the specific domain due to its open-source origin.  
  \item \textbf{GPT-4 Turbo}: In order to compare TP-GPT with a model that is not limited by model size, we conducted experiments on the up-to-date GPT model, GPT-4 Turbo, which has the 128,000 token limit. This model is commonly regarded as the most powerful generative intelligence by the public for its extensive pre-trained knowledge in various domains, absolutely including transportation and SQL query programming. In the experiment, the fundamental prompt containing contextual information such as description of the expected role to perform, database setup, and table schema is input to the model's memory. The result shows that GPT-4 Turbo achieves a better performance than PaLM 2, especially regarding the lower non-functional rate. The strong reasoning and coding capabilities enable the model to generate more executable SQL queries. However, obtaining flawless answers with correctly queried data is still challenging for its low solve rate compared to TP-GPT. A fair portion of queries are runnable yet do not properly answer the input questions.
  \item \textbf{TP-GPT}: Our innovative system TP-GPT, integrating intelligence-boosting strategies to regular GPT-4 model without need of fine-tuning, achieves the highest performance among powerful models. In experiments, TP-GPT is capable of crafting lengthy SQL statements within minutes and also interpreting data to answer tricky questions that require deduction and derivation. For example, TP-GPT tackles the question "\textit{What is the difference in greenhouse gas emissions between weekdays and weekends in the last month?}" by estimating emissions based on Vehicle Miles of Travel (VMT) using a conceptual formula.
\end{itemize}

Several observations are captured during experiments that potentially account for the superior performance of TP-GPT over GPT-4 Turbo: 1) TP-GPT's multi-agent strategy, especially the quality agent, enhances the query correctness by checking syntax errors. However, GPT-4 Turbo suffers from frequently executing incorrect queries without error checking. It also struggles to recognize the correct column name of a table even when the schema has been input; 2) TP-GPT's implementation of Chain of Thought enables the model to revise the response in multiple rounds of iterative communication. On the contrary, GPT-4 Turbo merely processes data in a sequential flow without the review loop, which causes a lack of understanding of the database environment and content. It is unable to inspect if the query will successfully extract correct data, having the potential to obtain empty results; 3) As TP-GPT is prompted by few-shot examples consisting of template questions and queries, higher proficiency in generating certain SQL dialects (e.g., Microsoft SQL) is observed for the prior knowledge leaned from few-shot learning in consistent contexts. With the absence of this functional module, GPT-4 Turbo fluctuates its performance due to frequently caused errors converting date from string to timestamp in Microsoft SQL syntax.

\subsection{Ablation Study}
This section examines the impact of removing the crafted prompt, few-shot learning, and multi-agent strategy from the proposed TP-GPT on TransQuery. Questions in the benchmark are input to TP-GPT and its three variations, and each response is evaluated as non-functional, runnable but imperfect, or flawless as described in Section \ref{Experimental Setup}. The percentage of responses in each of these three categories among all responses is calculated for the models, of which the results are shown in Figure \ref{fig:ablation}. By comparing the flawless response rates, TP-GPT significantly outperforms the three variations, validating the effectiveness of incorporating each of modules embedded in TP-GPT.  
\begin{itemize}
    \item \textbf{Prompt}: The prompt elaborately customized for transportation-domain inquiry is essential to TP-GPT. The removal of prompt leads to a sudden drop in flawless response rate from 80\% to 26\%, due to a lack of contextual knowledge and performing role description.
    \item \textbf{Few-shot learning}: Few-shot learning has the least impact on TP-GPT among modules regarding response metrics. This is because it only affects answering questions highly similar to the exemplar. For most questions not relevant to example inventory, the responses are barely impacted.
    \item \textbf{Multi-agent strategy}: Multi-agent collaboration plays a vital role in TP-GPT, suggested by the decrease in flawless response rate to 44\% when it is removed. Intuitively, the problem-solving process has multiple stages, where reviewing outputs, generating feedback, and creating thoughts iteratively are fundamental to producing accurate results. However, a sequential process with no reflection loop in this variation frequently results in misunderstanding of contexts, causing syntax errors, or extracting mistaken data.
\end{itemize}

\begin{figure}[h]
    \centering
    \includegraphics[width=\columnwidth]{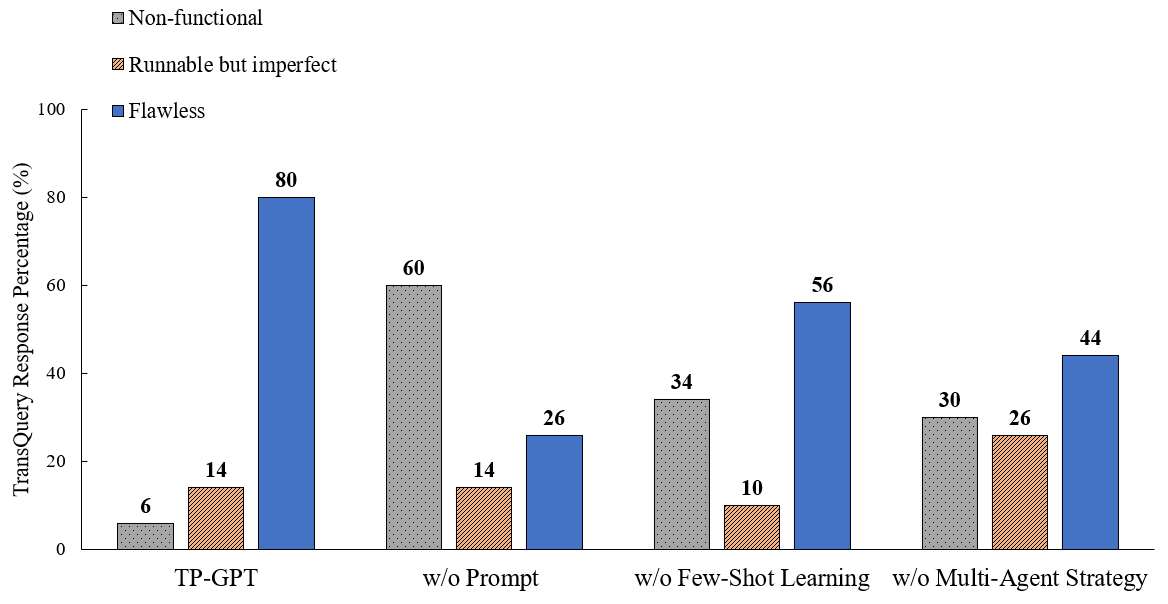}
    \caption{Ablation study of TP-GPT on TransQuery by removing prompts, few-shot learning, or multi-agent strategy}
    \label{fig:ablation}
\end{figure}

\section{Conclusions}
The integration of Large Language Models (LLMs) into the domain of traffic management presents a transformative approach to addressing the complexities and challenges inherent in modern transportation systems. This paper has outlined the development of an intelligent traffic performance chatbot, TP-GPT, which leverages the power of real-time traffic data and the contextual understanding of LLMs to provide efficient, accurate, and privacy-preserving transportation surveillance and management.

TP-GPT demonstrates a novel framework that can effectively navigate the vast and intricate traffic database landscapes to extract meaningful insights. It has great capabilities to understand contextual information and respond to inquiries in the transportation domain by transforming input texts into queries and converting data into detailed natural-language analysis reports leveraging extensive prior knowledge. TP-GPT employs Chain of Thought prompting for iterative query generation, a multi-agent strategy to optimize intermediate results, few-shot learning to enhance exemplar performance, and chat memory to improve interaction quality. The experimental study devices a challenging traffic-analysis benchmark TransQuery to compare the performance of TP-GPT with state-of-the-art baseline LLMs. Quantitative results show that TP-GPT significantly outperforms GPT-4 Turbo, PaLM 2, and SQLCoder, demonstrating its superior performance in resolving real-world transportation inquiry tasks. The ablation study validates the effectiveness of employed modules in TP-GPT. Furthermore, an intelligent online ChatBot empowered by TP-GPT is launched with an interactive web interface.

Future studies will improve TP-GPT to have the ability to predict future traffic conditions based on historical data, by possibly integrating traffic forecasting models. Besides, location recognition could be further enhanced leveraging the visual intelligence of language models in analyzing maps.

\bibliographystyle{ieeetr}
\bibliography{main}

\vfill 
\end{document}